\documentstyle[12pt,epsfig]
{article}

\def\gsim{\mathrel{\rlap {\raise.5ex\hbox{$ > $}}
{\lower.5ex\hbox{$\sim$}}}}
\def\lsim{\mathrel{\rlap {\raise.5ex\hbox{$ < $}}
{\lower.5ex\hbox{$\sim$}}}}

\newcommand{\be}{\begin{equation}}
\newcommand{\ee}{\end{equation}}
\newcommand{\bea}{\begin{eqnarray}}

\newcommand{\eea}{\end{eqnarray}}

\def\gappeq{\mathrel{\rlap {\raise.5ex\hbox{$>$}}
{\lower.5ex\hbox{$\sim$}}}}

\def\lappeq{\mathrel{\rlap{\raise.5ex\hbox{$<$}}
{\lower.5ex\hbox{$\sim$}}}}

\begin{document}

\begin{titlepage}

\begin{flushright}
DFPD-00/TH/39 \\
\end{flushright}

\vspace{0.1in}

\begin{center}
{\Large  Can   Neutrinos Probe   Extra Dimensions?}\\
\end{center}

\vspace{0.3in}

\begin{center}

 \vspace{.15in}
 {\large V. Ammosov$^{b}$
   $\,$and$\,$ G. Volkov$^{a,}$ $^{b}$\\}
 \vspace{.25in}
{\it  $^a$ Universita di Padova, Dipartimento di Fisica Galileo Galilei,\\
INFN, Padova , Italy.\\}
 {\it $^{b}$ Institute for High Energy Physics, Protvino, Russia.\\}
 \vspace{.05in}

\end{center}

\begin{abstract}
We conjecture that the topological structure  of the gauge symmetries
required by the Calabi-Yau vacuum and the dualities in string/D-branes
considered in the world with some additional
dimensions can lead to an extension
 of the main principles of the Special Theory of Relativity.
The link between the topological structure of the vacuum and the,
hierarchy of the gauge symmetries could be checked by the existence of
the ``almost massless'', ''sterile'' particles. These particles could have
flying properties different from the standard predictions of the STR.
It is natural to consider this property for neutrinos, known up to 
now ``$U(1)_{em}$
sterile'' particles.
Here we discuss  two such effects depending on the
possible existence of large and/or small
extra dimensions and
what is the maximal speed for both cases.

The  effect of large extra dimensions can be
directly  connected with the existence of a
hidden boost, $c_{hid} >> c_{em}$, excited by new global dimensions and can
lead to the monotonous rise of the neutrino velocity at high energies,
$v_{\nu} > c$.

The effect of small extra dimensions, universal with respect to all
particles, can be connected with
the gravitation recoils of the propagation of neutrino in the space-time
vacuum-foam and it leads to the effect of diminishing the neutrino velocity 
at high energies.

We propose to check these conjectures
for neutrinos of  different energies and species.
Limits for existing neutrino data and expected sensitivities of possible
experiments  for accelerator - produced neutrinos are considered. It is
pointed out that
CERN has the unique opportunity of measuring the indicated effects.

\end{abstract}

\end{titlepage}

 \section{\large  Strings dualities and the origin of the gauge symmetries
in the extra dimensional geometry.}
Our main experience after the studies  of the
dualities of string/D-branes \cite{Dual},\cite{Drev}
and the origin of the gauge symmetries \cite{Gauge} in
the topological vacuum
  showed a very intriguing consequence for the
geometry(extra dimensions and topology)  of our ambient space-time.
A topological structure of the vacuum allows to  prove  some dual
transformations that interrelate any of
the five superstring 10-dimensional theories, the
M/11-dimensional supergravity and 12-dimensional F-theory with each other
\cite{MFrev}.
 The proof the dual relationships
among  all of these  theories can be done  in the process
of a compactification (or decompactification) on (of) the special set of
these topological
hypersurfaces with its very beautiful geometric topological
substructure. The most intriguing inspiration of 
 proving  the
string/D-brane dualities is that the ``singularities''
of these hypersurfaces are connected with the mechanism of the
enhanced gauge symmetries and give much more deeper understanding of
the  origin of the gauge and chiral matter \cite{Gauge}.

From the point of view of the topological quantum geometry the question of
the
compactification or/and decompactification has  been already
studied in the literature intensively (see, for example, \cite{Gre}).
From the phenomenological point of view there
was only the question of the sizes of these new topological objects.
Following to our goal to understand what kind of global real effects
could lead to the existence of the extra topological geometry
it will be important to consider separately two cases
for the possible sizes of the new extra dimensions:
the small (compactified)  extra dimensions with the sizes determined
by the scale from Plank scale till the scale of the present energies
and the large (uncompactified)
extra dimensions which size can reach even  infinity.

The first case of the possible existence of the small (compactified) extra
dimensions
has been intensively studied in the literature, see for example \cite{Min}.
In  the case of uncompactified extra dimensions,
\cite{Rand}
 the Newton law of gravity should be overcome and this possibility
can give  a lot of important cosmological consequences.
In this case our world can be  considered, for example,
as a ``border-part'' of the higher
dimensional world  with the global uncompactified 5-th or 6-th dimensions
and with other 6- compactified dimensions. It is interesting to note that
both cases can occur in M/F theories  with {dim=11/12}.

An interesting example of the possible realisation
of such case was found in \cite{Mald} where it was shown that there is
 duality between the Yang-Mills N=4 SUSY theory
in 4-dimensions and the supergravity in 5-dimensional anti-de Sitter space.
In string/D-brane approach our four-dimensional SM world could live
on the world volume of a 3-brane with flat topology, which is embedded
in a bulk space-time of dimension, $D > 4$. In this case the metrics of our
world is induced by the metrics of the higher dimensional bulk space-time.

Of course, in the process of the extension of such world
there could be some interrelations between the compactified and
un-compactified dimensions due to shrink-down or decompactification.
The process of
compactification of some topological manifolds can be the origin  of the
Yang-Mills gauge symmetries and matter.
The new uncompactified dimensions could lead to some new phenomena,
to the possibility of the existence of a new matter with the different
global properties with respect to the new space-time symmetries with
possible
new boosts.
We already had one experience after the discovery the Lorentz invariance
of the Maxwell equations which gave a possibility to understand the ambient
Minkowski space-time with global $SO(3,1)$ symmetry.
Kaluza and Klein understood that the unification of the $U(1)_{em}$
gauge symmetry and four-dimensional gravity can be naturally
explained in the frame of the five-dimensional gravity
where the new compactified dimension has a nontrivial topology.
This link between the internal and
external symmetries of vacuum, Poincare duality, could be now considered
again
after the discovery of  the new vacuum internal symmetry,
$SU(2) \times U(1)$, at the smaller distances.
 In this case due to Poincare duality there could be a link between
the hierarchy of the gauge symmetries and new geometry of the space-time
with the possible new dimensions and its topological structure.

There  exists a very intriguing question
how to check the Poincare duality and to prove the existence of  these
new extra dimensions. We propose
to study this question through the searches of the new space-time
global effects at high energies.
The development of string/D-branes approach and the existence at high
energies of the new vacuum symmetry, $SU(2)\times U(1)$, give us a chance to
move forward
in understanding an origin of the speed fundamental constant.
Our main hope is connected with the
possibilities to find a deviation for
the principle of the light speed maximality , $c=c_{em}$,
studying the flight properties of the high energy neutrinos
comparing it velocity with the speed of light or with the
speeds of the other charged relativistic particles.

Our proposal consists on the   possibilities to check at the present 
accelerators the existence
of the large and small extra  dimensions measuring
 the neutrino speed at the high energy in the different scales and
for different species. We conjecture that the global large (uncompactified)
 extra and small (compactified) dimensions
 could lead to the observation of
the two different behaviours of the neutrino speed
at high energies.

\section{\large  Neutrinos  and large extra dimensions}

We conjecture that the large new dimensions could lead directly to the
observation of
the effect of the monotonous rise with the energy of the neutrino velocity ,
$v_{\nu}(E\sim E_{thr})>c_{em}$ 
($E_{thr}$ is the  energy of the $SU(2)\times U(1)$ symmetry restoring), 
  and   could be searched by
with "$U(1)_{em}$-sterile" neutrinos of high energies.
In addition to the Poincare duality, having very nice interpretation
in algebraic geometry,  some new arguments for the possible
existence of the new  symmetry with the new boost,  $c_{hid}>>c_{em}$,
could be found from  the considerations together of the modern achievements
in
neutrino physics, in astrophysics and cosmology, in phenomenology of
of the dualities in string/D-branes, in algebraic geometry of the
topological hypersurfaces and its link with the origin of the gauge symmetry
and matter .
So,  one can propose that this possibility could be realized with the
following conditions:

\begin{itemize}
\item{ The existence of a new ambient geometry  with new hypothetic
"sterile" world
connected with a possibility of the present or previous  existence
of a new extra dimension}\\
\item{ The sterile world should satisfy the own space-time
symmetries incorporated a  new boost $c_{hid}>>c_{em}$;}\\
\item{ There should exist some gauge symmetries of the sterile matter
which could be linked with the observable SM -matter through neutrinos of
the different species. The mechanism of the mixing between the
sterile matter and  neutrinos could be similar to the
``sea-saw'' mechanism.}\\
\end{itemize}

So, we conjecture that the duality-symmetry between the topological
structure of the vacuum  and
new hierarchy of the gauge symmetries:

\begin{eqnarray}
U(1)_{em} \Longrightarrow SU(2) \times U(1) \Longrightarrow ....
\end{eqnarray}
should lead to an extension of the special theory
of relativity based on the ``one-circuit'' Lorentz metrics
$g_{mn}=(-1,-1,-1,1)$
to the new ``two-circuit'' metrics with
the possible existence of the new boost:

\begin{eqnarray}
\{ SO(3,1) \}(c_{em})\,\, \Longrightarrow \,\,
\{ G^{ext}\subset SO(3,1) \times SO(1,1)\times ... \}
(c_{em},\,c_{hid})\,\,\Longrightarrow\,.....
\end{eqnarray}

Thus an extension of the Lorentz/Poincare symmetry at high energies
could lead to the existence of the new hidden  symmetries,
like hidden boosts with the new fundamental maximal speeds, $c_{hid}$,
($...>>c_{hid} >> c_{em} $), which
could be checked by the existence of the `` $U(1)_{em}$- sterile''
particles.
Based on the Poincare duality between the external symmetries of the ambient
geometry and the internal gauge symmetry of the vacuum  we propose
 the existence of the new hidden space-time symmetry- new hidden boost-
(not necessarily the light )
 excited be the extra dimensions of our world
connected with the $ SU(2) \times U(1) $ vacuum symmetry.
This new boost could appear, for example, from the breaking of the
extra world space symmetries, like subgroup of $SO(n,1)$ or $SO(n,2)$.

The question of the existence of the
``sterile-hidden'' matter has been already intensively discussed
in astrophysics and cosmology during a long period. In the first condition
one had suggested that this matter can produce a 
hidden new geometrical world including new extra dimensions
with additional  space-time symmetries. Of course, by this scenario the
hidden matter  should be concentrated mainly on the very long distances.
In this case the existence of this additional symmetries should include
a hidden boost with $c_{hid}>>c_{em}$ according to our second condition.
From our knowledge of the SM the neutrinos are unique and  the only
particles which could have a link with the 'sterile' world.
One might propose that the origin of 
the "see-saw" mechanism
could also explain the link between our neutrinos and the sterile particles.
Due to mixing between neutrino and sterile matter through see-saw mechanism
the global flying properties of neutrino produced from the decays of
the sterile matter should correspond to the space-time symmetry of the extra
world.
 The sterile matter world should have a  new mass scale
higher than the electroweak scale.

As result of this link  our neutrinos could ``feel''
the new boost, $c_{hid}$, and  could lead to the
increasing of the maximal velocity for neutrinos with restoring of the
$SU(2)\times U(1)$-vacuum symmetry.

The model
with one new hypothetical boost
could be realized also through the two-metrics mechanism:
 one-metric action, $S_{SM}$,
for the Standard Model based on the Lorentz/Poincare\, symmetry
in the $d= 3+1$ space-time and
the other action, $S_{hid}$,  for the new sterile matter based
on the new metric tensor in the space-time with
extra dimensions.
For the concrete model one should solve the main difficulty
of finding a link between these two metrics.

The effective metrics near the threshold of the $SU(2) \times U(1)$
symmetric vacuum, deviated from the standard dispersion relation between
the energy {\it E} and the momentum {\it p} of neutrino, could be a function
of the invariant production energy {\it s}, the ``wind'' energy {\it E}
and the relation between the values of two circuit-boosts, $c_{em}$ and
$c_{hid}$ and, may be, also depend on the  neutrino species.

Such scenario with some fundamental boost-velocities
could give a push to go beyond the standard Big Bang model
in the time before the $SU(2)\times U(1)$ phase and  could  explain
the horizon, flatness  problems. This scenario is different with respect
to
the other scenario of the varying speed of light (VSL)
\cite{Moffat} in spite of  the similar
 problems  to construct such mechanisms.

The restriction  of sterility for neutrino in our conjecture means
that with the electromagnetic charged particles
to observe new boost  (new topological circuit) at the now 
available  energies is very difficult
or may be impossible now. By our scenario only neutrinos could be link
with the sterile hidden matter action and could ``feel'' the
second boost-speed.
For $ U(1)_{em}$ charged particles ``getting'' to the new
$SU(2) \times U(1)$ vacuum structure could be a threshold effect like as
Vavilov-Cherenkov effect with emitting of a lot of energy.

\section{\large Neutrinos and small extra dimensions}
The other possibility is connected with the new compactified
small dimensions.
It will be interesting  to compare the experimental possibility of
searching for the
 effects, $v_{\nu}(E)> c_{em}$, for neutrinos with the so called
EMN effect \cite{Ell} leading to the diminishing of the particle
velocity  $< c_{em}$ with increasing  energy.
This effect  is connected with the string/D-brane gravitation foam structure
of the vacuum and the magnitude of this effect depends on the
D/brane mass, M. It is   universal for all particles  and leads
to the diminishing of the velocity of this particle with increasing
of energy:
\begin{equation}
c(E)\,<\,c_{em}
\end{equation}
To understand this  point  one can follow to the way
suggested in papers
\cite{Ell} where energy dependence of the ``effective'' metric  is the
main deviation from space-time Lorentz invariance
induced by the $D$ particle recoil.
It has been argued that  virtual $D$-
branes provide one possible model for space-time foam~\cite{Ell}, and
that the recoil of a $D$ brane struck by a bosonic/fermionic
closed-string particle would induce an energy-dependent
modification of the background metric.
Upon diagonalization of the perturbed metric, one finds
a retardation in the propagation of
an energetic photon/fermion : $(\delta c / c) \sim (E / M)$
or $(\delta c / c) \sim (E /{\tilde  M})^2$.
One can also see that the correct dispersion relation between the
energy $E$ and momentum $p$ of
the massless particle in the metric background \cite{Ell} is:
\begin{equation}
{E}^2\,=\,c^2\cdot {p}^2\,-\,2 \cdot c\cdot  E\cdot (\vec p \cdot \vec u):
\qquad
| \vec u |~ \sim E / M \,\,\,  or \,\,\,\sim (E / \tilde M)^2.
\label{dispferm}
\end{equation}

It has recently been pointed out that the constancy of $c$, the
velocity of light, can be tested stringently using
distant astrophysical sources that emit pulses of radiation, such
as Gamma Ray Bursts (GRBs), Active Galactic Nuclei (AGNs) and
pulsars. So far, this idea has been explored by comparing the
arrival times of photons of different energies $E$ (frequencies $\nu$).
It has been suggested that certain quantum theories of gravity might
cause variations in $c$ that increase with $E$ (or $\nu$), possibly
linearly: $(\delta c/ c) \sim (E / M)$, or quadratically:
$(\delta c/ c) \sim (E^2 / {\tilde M}^2)$, where $M$ or $\tilde M$
is a high mass scale characterizing quantum fluctuations in
space-time foam. Such a linear or quadratic dependence would enable
any such conjectured quantum-gravity effects to be distinguished
easily (in principle) from the effects of conventional media on
photon propagation and the effects of a possible photon mass,
both of which would {decrease} with increasing energy. It is clear that
in order to probe quantum-gravity effects by putting the strongest
possible lower limits on $M$ and $\tilde M$, there is a
premium on distant pulsed sources that emit quanta at the highest
available energy.

 In this approach the mass of
D- particle could be  large, $M_D >> M_W$, but nevertheless  the
correspondent
reduction of the effective Lorentz metrics can be checked for neutrinos too.
The example of this metrics could be important to verify
the similarly ideas with the scale $M, \tilde M$ not
very far from the 1-Tev scale.

\section{\large  Experimental tests. }

Awaiting for  further theoretical developments of  the basic structure
of our World it is needed to put these conjectures under  experimental
checks.

 In the past, the  idea to measure the  velocity for accelerator-produced
neutrinos
was  stated in  \cite{Kalb} and  the subsequent  experimental measurements
were performed  25 years  ago at the  Fermilab \cite{Als}.
The  following  upper  limits at  95\%  CL  were established

\centerline{${
| \Delta \beta_{12}| =
|v_{\nu_k}/c - v_{\nu_\pi}/c| < 5\cdot 10^{-5}}$,}

\centerline{${
|\beta_{\nu} - \beta_{\mu}|=|\Delta \beta | < 4\cdot 10^{-5}}$,}
where $\Delta \beta = v_{\nu}/c-1$ .

The  idea was  refreshed in  1987  \cite{Sto} when  the SN87A
explosion  was observed  using photon  \cite{She} and  neutrino
\cite{Hir} detections. The estimated upper limit is

\centerline{${ |\Delta \beta| <  10^{-8}}$.}

Furthermore the SN87A data can be used for estimates of differences
of neutrino velocities too.

Because  the 12.4  s experimental
duration of the neutrino pulse\cite{Dad} is compatible with expected
duration of
the  neutrino emission  during the  SN87A gravitation  collapse  it is
reasonable to take this value as an upper limit for the $\Delta \beta_{12}$
estimate with the average neutrino energy  $\sim 10 MeV$:

\centerline  {$\Delta  \beta_{12} <~\Delta T/T_0  =
12.4/4.5\cdot 10^{12} = 2.8\cdot 10^{-12}$,}

where $T_0$ = 150000 light years.

It seems that accelerator-produced high energy neutrinos, where  the
flavor  and the  source of  neutrinos are controlled,
can play  their own  important role.
Among other accelerator sites CERN has the unique  opportunity
with the short baseline, the long baseline and the beam-beam LHC
 neutrino experiments
 to  perform systematical measurements of neutrino velocities
using muon, pion and kaon, charm and beauty, W and Z decays.

The Table~ \ref{tab1} summarizes  existing results and expected sensitivity
estimations of possible experiments
for deviation of neutrino velocity from the speed of light .
As one can see from the Table~\ref{tab1}

\begin{itemize}
\item {WBB and beam-dump experiments will improve the Fermilab limit for
$\nu_{\mu}$ from pion and kaon decays
and beam-dump experiment can establish new limits  for
$\nu_{\mu}$ and  $\nu_{e}$ from charm decays;}\\
\item{ long baseline experiments will improve the
SN87A limit for $\nu_{e}$ and the Fermilab limit  for $\nu_{\mu}$;}\\
\item{ even beam-beam LHC experiment is feasible and will establish new
limits
 for $\nu_{\mu}$ ,  $\nu_{e}$ and $\nu_{\tau}$ from beauty, W and Z
decays.}\\
\end{itemize}

\begin{table}[th]
\begin{center}
\caption{Existing results and expected one standard
deviation estimations for neutrino velocity measurements
for the $\Delta {\beta}_{12}$
in the case of $v_{\nu}\approx c$ for one year running 
($ 10^7~ $ sec) assuming the
$\sigma = 250$ psec beam bunch filling and the
$\sigma = 200$ psec detector time resolution.}
\label{tab1}
\begin{tabular}{|c|c|c|c|c|}\hline \hline
Experiment                                  &  Source of
& Type of                                   &  Statistics
& Sensitivity                                             \\
$                         $                 &  neutrino
& neutrino                                  &  CC events
&$  of~ \Delta \beta      $                               \\  \hline\hline
FNAL, NBB CCFR$ ^{\star )}$ \cite{Als}      &  $ h=\pi +k $
&$  \nu_{\mu}             $                 &  $          $
&$4\cdot 10^{-5}          $                                \\ \hline
SN87A$ ^{\star )}         $\cite{Dad}       &  $          $
&$  \nu_{e}               $                 &  $          $
&$1\cdot 10^{-8}          $                               \\
\hline\hline
WBB  at the WANF\cite{Wanf}                 &  $ h=\pi +k $
&$  \nu_{\mu}             $                 &  $ 400000   $
&$1.3\cdot 10^{-6}        $                               \\
$2~t$ detector, $10^{19}  $ pot             &  $          $
&$ \nu_{e}                $                 &  $ 4000     $
&$ 2.2\cdot 10^{-6}       $                               \\ \hline
Beam-dump at the WANF                       &  $ h=\pi +k $
&$  \nu_{\mu}             $                 &  $ 47000    $
&$ 2\cdot 10^{-6}         $                               \\
$ 2~kt$ detector,$10^{19} $ pot             &  $          $
&$ \nu_{e}                $                 &  $ 500      $
&$ -                      $                               \\ \cline{2-5}
$                         $                 &    charm
&$  \nu_{\mu}             $                 &  $ 46000    $
&$1.5\cdot 10^{-6}        $                               \\
$                         $                 &  $          $
&$ \nu_{e}                $                 &  $ 46000    $
&$1\cdot 10^{-6}          $                               \\  \hline
beam-beam at the LHC                        &    charm
&$  \nu_{\mu}             $                 &  $ 200000   $
&$2.5\cdot 10^{-6}        $                               \\
$50~kt$ detector for $L=10^{35}$\cite{Gar}  &  $          $
&$ \nu_{e}                $                 &  $ 200000   $
&$2.5\cdot 10^{-6}        $                               \\ \cline{2-5}
$500~m                    $ flight pass     &     beauty
&$  \nu_{\mu}             $                 &  $  30000   $
&$2.5\cdot 10^{-6}        $                               \\
&$                        $
&$ \nu_{e}                $                 &  $  30000   $
&$2.5\cdot 10^{-6}        $                               \\ \cline{2-5}
$                         $                 &  $ W+Z      $
&$  \nu_{\mu}             $                 &  $ 230      $
&$1.7\cdot 10^{-5}        $                               \\
$                         $                 &  $          $
&$ \nu_{e}                $                 &  $ 230      $
&$ 1.7\cdot 10^{-5}       $                               \\
$                         $                 &  $          $
&$ \nu_{\tau}             $                 &  $ 230      $
& $1.7\cdot 10^{-5}$                                      \\ \hline
Long baseline at the NGS\cite{NGS}          &  $ h=\pi +k $
&$  \nu_{\mu}             $                 &  $ 5000     $
&$1.8 \cdot 10^{-9}       $                               \\
$10~kt$ detector, $10^{19}$ pot             &  $          $
&$ \nu_{e}                $                 &  $  50      $
&$1.8 \cdot 10^{-8}       $                               \\  \hline
LBL at the NuFact\cite{Nuf}                 &  $ \mu      $
&$  \nu_{\mu}             $                 &  $ 14000    $
&$3 \cdot 10^{-10}        $                               \\
$10~kt$ detector,3000 km, $10^{21}\mu $     &  $          $
&$ \nu_{e}                $                 &  $ 14000 ,  $
&$3 \cdot 10^{-10}        $                               \\  \hline\hline
\end{tabular}
\end{center}
{~~~~$*)$ existing result}
\end{table}
\normalsize

Below in the Table~\ref{tab2} we compare sensitivities of the EMN
effect for neutrino velocity differences
for existing and possible experiments for $\delta \beta = E/M$
 and $\delta \beta = E^2/\tilde M^2$ hypothesis.

\begin{table}[th]
\begin{center}

\caption{Existing results and expected sensitivities
for  $\Delta \beta_{12}$  neutrino velocity differences.
Expected values are given for one year running ($10^{7}~$sec) assuming
the $\sigma =250~psec$ beam bunch filling and the
$\sigma=200~psec$ detector time resolution.}
\label{tab2}

\begin{tabular}{|l|c|c|c|c|c|}\hline\hline
Item                        &$E_{\nu}         $,MeV&$E_{\nu}          $, MeV 
&$\Delta \beta_{12}     $   &$M,\, GeV        $    &$ \tilde M, \,GeV $ \\
                            & up point             &  low point           
&                           &$                $    &$                 $ \\
\hline\hline
SN87A, KMK$ ^{\star )}  $   & 15                   &  7 
&$<~2.8\cdot 10^{-12}   $   &$>~2.8\cdot 10^9 $    &$ >~8\cdot 10^{3} $ \\
\hline
FNAL, CCFR$ ^{\star )}  $   &$1.7\cdot 10^5   $    &$ 5\cdot 10^4     $ 
&$<5\cdot        10^{-5}$   &$>2.5\cdot 10^{6}$    &$ >2.4\cdot 10^{4}$ \\ 
\hline\hline
WBB at WANF                 &$10^{5}          $    &$ 2\cdot 10^4     $ 
&$ 1.8\cdot 10^{-6}     $   &$4.2\cdot 10^{7} $    &$ 7\cdot 10^{4}   $ \\
\hline
BD at WANF                  &$10^{5}          $    &$ 2\cdot 10^4     $ 
&$1.4\cdot 10^{-6}      $   &$5.5\cdot 10^{7} $    &$ 8\cdot 10^{4}   $ \\
\hline
B-b at LHC                  &$2\cdot 10^{6}   $    &$ 2 \cdot 10^4    $ 
&$1\cdot 10^{-5}        $   &$2\cdot 10^{8}   $    &$ 6\cdot 10^{5}   $ \\
\hline
LBL at NGS                  &$3\cdot 10^{4}   $    &$ 10^4            $ 
&$3.6\cdot 10^{-9}      $   &$5.5\cdot10^{9}  $    &$ 5\cdot 10^{5}   $ \\
\hline
NuFact                      &$5\cdot 10^{4}   $    &$ 2\cdot 10^4     $ 
&$4.2\cdot 10^{-10}     $   &$7.1\cdot 10^{10}$    &$ 2.2\cdot 10^{6} $ \\
\hline\hline
\end{tabular}
\end{center}
{~~~~$*)$ existing result}
\end{table}
\normalsize

As one can see  the long baseline
experiments can  exceed the  SN87A low limit  for the  $\delta \beta
= E/M$  hypothesis  and  all  accelerator  experiments  are  more
sensitive  than  the  SN87A  for  the  $\delta  \beta  =E^2/\tilde M^2$
hypothesis.

As it was pointed out in section 2 the sterile matter should be heavy
and therefore it could be the  source of the very high energy
neutrinos  from  far ends of our ``visible''
Universe. If , for example, such astrophysical objects like GRBs
or/and AGNs  satisfy the conditions for the sterile matter
creation then
they could be a source for neutrinos with velocities exceeding the speed
 of light.
Due to huge distances from these objects
the sensitivity  $\Delta \beta \sim 10^{-16}$  can be reached \cite{Wax}.

\section{\large Acknowledgements }
We would like to thank F. Anselmo, F. Dydak, L. Fellin,
E. Gousctchin, G. Harigel, J. Ellis,
 D. Nanopoulos, V. Petrov, M. Pietroni, T. Ypsilantis and A. Zichichi 
for many valuable discussions and support.
We would like to thank G. Costa for a careful reading of the manuscript and
many useful comments. 
One of us (VA) thanks G. Gapienko for help with estimates.
 (GV) thanks also for the hospitality the Department of Physics 
``Galileo Galilei'' of the Padova University and the Padova Section of INFN,
where part of this work was carried out.
\newpage
~

\end{document}